\documentclass[aps,twocolumn,prb,superscriptaddress,floats,nobibnotes,notitlepage,citeautoscript]{revtex4-2}

\usepackage{graphicx}
\usepackage[utf8]{inputenc}
\usepackage{amsmath,amssymb}
\usepackage{newtxtext}
\usepackage[varvw]{newtxmath}
\usepackage{multirow}
\usepackage{mathtools}
\usepackage[pdftex]{hyperref}

\usepackage{dsfont}
\usepackage{bbold}
\usepackage{color,soul}
\usepackage{xstring}
\usepackage{wasysym}
\usepackage{xspace}
\usepackage{time}
\usepackage{bbold}
\usepackage{subfigure}
\usepackage{multirow}
\usepackage{xspace}
\usepackage{url}
\usepackage{xcolor}

\hypersetup{
    colorlinks,
    linkcolor={blue},
    citecolor={blue},
    urlcolor={blue}
}

\def\<{\langle}
\def\>{\rangle}
\DeclareMathOperator{\Tr}{Tr}
\usepackage[normalem]{ulem}

\newcommand{\ve}[1]{\mathbf{#1}}
\newcommand{\rmi}{\mathrm{i}}
\newcommand{\rme}{\mathrm{e}}
\newcommand{\rmd}{\mathrm{d}}

\date{\today}

\begin{document}

\title{Critical properties of metallic and deconfined quantum phase transitions in Dirac systems}

\author{Zi Hong Liu}
\affiliation{Institut f\"ur Theoretische Physik and W\"urzburg-Dresden Cluster of Excellence ct.qmat,
Technische Universit\"at Dresden, 01062 Dresden, Germany}
\author{Matthias Vojta}
\affiliation{Institut f\"ur Theoretische Physik and W\"urzburg-Dresden Cluster of Excellence ct.qmat,
Technische Universit\"at Dresden, 01062 Dresden, Germany}
\author{Fakher F. Assaad}
\affiliation{Institut f\"ur Theoretische Physik und Astrophysik and W\"urzburg-Dresden Cluster of Excellence ct.qmat,\\
Universit\"at W\"urzburg, 97074 W\"urzburg, Germany}
\author{Lukas Janssen}
\affiliation{Institut f\"ur Theoretische Physik and W\"urzburg-Dresden Cluster of Excellence ct.qmat,
Technische Universit\"at Dresden, 01062 Dresden, Germany}

\begin{abstract}
We characterize, by means of large-scale fermion quantum Monte Carlo simulations, metallic and deconfined quantum phase transitions in a bilayer honeycomb model in terms of their quantum critical and finite-temperature properties.
The model features three different phases at zero temperature as function of interaction strength. At weak interaction, a fully symmetric Dirac semimetal state is realized. At intermediate and strong interaction, respectively, two long-range-ordered phases that break different symmetries are stabilized. The ordered phases feature partial and full, respectively, gap openings in the fermion spectrum. 
We clarify the symmetries of the different zero-temperature phases and the symmetry breaking patterns across the two quantum phase transitions between them.
The first transition between the disordered and long-range-ordered semimetallic phases has previously been argued to be described by the $(2+1)$-dimensional Gross-Neveu-SO(3) field theory.
By performing simulations with an improved symmetric Trotter decomposition, we further substantiate this claim by computing the critical exponents $1/\nu$, $\eta_\phi$, and $\eta_\psi$, which turn out to be consistent with the field-theoretical expectation within numerical and analytical uncertainties.
The second transition between the two long-range-ordered phases has previously been proposed as a possible instance of a metallic deconfined quantum critical point. We further develop this scenario by 
analyzing the spectral functions in the single-particle, particle-hole, and particle-particle channels. Our results indicate gapless excitations with a unique velocity, supporting the emergence of Lorentz symmetry at criticality.
We also compute the finite-temperature phase boundaries of the ordered states above the fully gapped state at large interaction. 
The phase boundaries vanishes smoothly in the vicinity of the putative metallic deconfined quantum critical point, in agreement with the expectation for a continuous or weakly-first-order transition.
\end{abstract}

\maketitle

\section{Introduction}
\label{sec:intro}

Quantum critical points refer to continuous phase transitions occurring at absolute zero temperature. In two-dimensional systems consisting of only few atomically-thin layers, a number of control parameters that may potentially drive such transitions exist, such as uniaxial or hydrostatic pressure, lattice mismatch, or twisting angle.
In insulators, conventional quantum critical points are characterized by fluctuations of a bosonic order parameter alone. These quantum critical points can usually be fully understood in terms of a corresponding higher-dimensional classical transition within the Landau-Ginzburg-Wilson paradigm. 
In metals, by contrast, the presence of gapless fermionic degrees of freedom at the transition inhibits such a quantum-to-classical mapping.
As potential platforms for physics beyond the Landau-Ginzburg-Wilson paradigm, metallic quantum critical points have therefore attracted significant attention in recent years~\cite{metlitski10a, metlitski10b, sur16, schlief17, lee18, berg19, klein20, esterlis21, guo22, liu23, frank23}.
A different route towards the exploration of unconventional phase transitions has been the physics frustrated quantum magnets~\cite{vojta18}. In the presence of significant quantum fluctuations arising from frustration, the system can feature fractionalized excitations, interacting via emergent gauge fields.
Such a scenario has been heavily discussed at quantum phase transitions between N\'eel antiferromagnetic and valence bond solid phases in magnetic Mott insulators~\cite{senthil04, kuklov08, kaul13, nahum15, shao16, zhao20, cui23, zhao22, song23, deng24, demidio24}.
Since the two states adjacent to the phase transition point break different symmetries, a continuous and direct transition is, without fine tuning, not possible if the transition is governed by fluctuations of the order parameters alone.
The numerical evidence for a generic continuous (or weakly first-order) transition has therefore been interpreted as a consequence of the emergence of fractionalized quasiparticle excitations that are confined in both long-range-ordered phases, but become deconfined at the transition point~\cite{senthil23}.
A related class of unconventional quantum critical points has been discussed in models that feature transitions into phases characterized by topological order, such as quantum spin liquids. If the fractionalized excitations associated with the topological order become or remain gapless at the transition point, they give rise to exotic fractionalized quantum universality classes that do not have any classical analogues~\cite{isakov12, schuler16, gazit18, janssen17, janssen20, seifert20, borla24}.

In this work, we study a lattice model that features two metallic quantum phase transitions as function of coupling strength~\cite{liu22}. The first one is continuous and can be understood within the $(2+1)$-dimensional Gross-Neveu-SO(3) [GN-SO(3)] field theory~\cite{ray21}. In the  SO(3)-broken phase, two out of  three  Dirac cones  acquire a mass,  and  the wave  function of the ungapped  Dirac  electron  couples  to  the SO(3)  order parameter.    Similar   mass terms  have been put  forward in spin-orbit-coupled fermions on a honeycomb lattice~\cite{Mondal23}. 
The second one is a transition between two long-range-ordered phases that break different symmetries: the aforementioned SO(3)-spin-symmetry-broken semimetal and a U(1)-charge-symmetry-broken fully gapped state~%
\footnote{Note that the ``fully gapped state'' exhibits a finite gap in the single-particle spectrum while still retaining a gapless Goldstone excitation in the two-particle spectrum, associated with the spontaneously broken U(1) symmetry.}.
The latter can be understood as an insulator characterized by U(1) interlayer coherence~\cite{liu22}. We show, nevertheless, that the interlayer-coherent insulator is degenerate with an $s$-wave superconducting state as a consequence of a partial particle-hole (PPH) symmetry.
At the level of mean-field theory, the transition between the SO(3)-spin- and U(1)-charge-symmetry-broken phases is strongly first order~\cite{liu22}.
Quantum fluctuations beyond mean-field theory significantly weaken this order-to-order transition, rendering it a candidate for a deconfined quantum critical point in the presence of gapless fermionic excitations---a potential microscopic realization of metallic deconfined quantum criticality~\cite{zou20}.
We use large-scale sign-problem free quantum Monte Carlo (QMC) simulations with a Hermitian Trotter decomposition, which significantly improves convergence properties with respect to the limit of small Trotter time steps, to compute the quantum critical and thermodynamic properties of both transitions.
Furthermore, in contrast to our initial exploratory study of this model~\cite{liu22}, we employ a microscopic implementation that preserves the model's PPH symmetry explicitly, maintaining the degeneracy of the interlayer-coherent insulating and $s$-wave superconducting ground states in the fully gapped phase already on finite lattices.
The interlayer-coherent insulating and $s$-wave superconducting states allow finite-temperature phases above the zero-temperature order.
The latter breaks both U(1)-charge and PPH symmetries. 
Upon increasing temperature, it is therefore possible that the U(1) and PPH orders melt at different temperatures, with an intermediate vestigial phase at intermediate temperatures~\cite{svistunov15, fernandes19, francini24}.
We compute the finite-temperature phase diagram of the model and show that the boundaries of the low-temperature orders vanish upon approaching the putative deconfined quantum critical point, in agreement with the expectation for a continuous or weak-first-order transition.

The rest of this work is organized as follows: 
In Sec.~\ref{sec:model}, we introduce our model. Its symmetries are discussed in Sec.~\ref{sec:symmetries}.
Details of our QMC simulations, with a focus on the technical advances achieved in the present work in comparison with our initial exploratory study of the model~\cite{liu22}, are given in Sec.~\ref{sec:QMC}.
Section~\ref{sec:zerotemperature} contains a discussion of our results on the critical properties of the two different quantum phase transitions occurring in our model at zero temperature. 
In Sec.~\ref{sec:thermodynamics}, we present our results on the finite-temperature phase diagram of the model.
We conclude with a summary and outlook in Sec.~\ref{sec:conclusions}.

\section{Model}
\label{sec:model}

\begin{figure}[tb]
\begin{centering}
\includegraphics[width=\columnwidth]{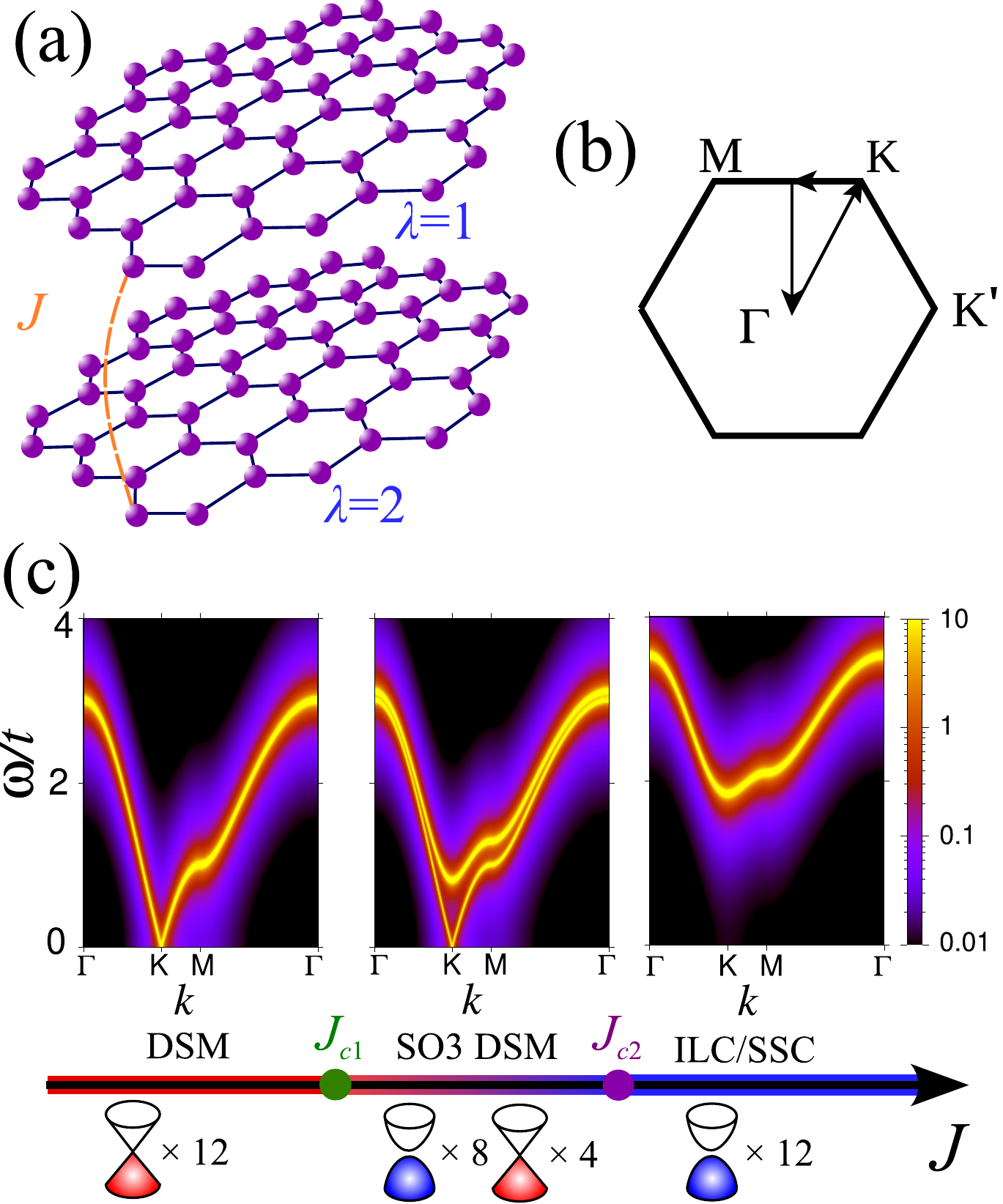}
\par\end{centering}
\caption{%
(a)~Sketch of bilayer honeycomb model defined by the Hamiltonian given in Eqs.~(\ref{eq:model1})-(\ref{eq:model3}). Each lattice site (purple dots) contain three SO(3) spin flavor degrees of freedom. The two layers interact via an SO(3)-symmetric spin-density interaction parametrized by the coupling $J$ (orange link).
(c)~Bottom: Schematic quantum phase diagram as function of $J$, showing the disordered Dirac semimetal (DSM) at small $J$, the SO(3)-spin-symmetry-broken Dirac semimetal (SO3 DSM), in which $2/3$ of the Dirac cones acquire a mass gap, while $1/3$ remains gapless, at intermediate $J$, as well as the fully gapped degenerate interlayer-coherent insulator (ILC) and $s$-wave superconductor (SSC) at strong $J$. The insets illustrate the fermion band structures in the three different phases.
Top: Fermion single-particle spectral function from lattice mean-field theory along a high-symmetry path in the first Brillouin zone [depicted in (b)], illustrating the partial and full gap openings in the SO3 DSM and ILC/SSC phases, respectively.
\label{fig:model}}
\end{figure}

We consider a model of interacting complex fermions on the bilayer honeycomb lattice, defined by the Hamiltonian
\begin{align} \label{eq:model1}
\hat{H} & =\hat{H}_{0}+\hat{H}_J
\end{align}
with nearest-neighbor intralayer hopping part
\begin{align} \label{eq:model2}
\hat{H_0} & = -t\sum_{\left\langle ij\right\rangle}\hat{c}_{i,\sigma,\lambda}^{\dagger}\hat{c}^{}_{j,\sigma,\lambda}
\end{align}
with hopping parameter $t>0$ and an SO(3)-symmetric on-site interaction part
\begin{align} \label{eq:model3}
\hat{H}_J & = -J\sum_{i,\alpha} \left(\hat{c}_{i,\sigma,\lambda}^{\dagger}K_{\sigma\sigma^{\prime}}^{\alpha}\tau_{\lambda\lambda^{\prime}}^{z}\hat{c}^{}_{i,\sigma^{\prime},\lambda^{\prime}}\right)^{2}.
\end{align}
with coupling $J \geq 0$.
Here, $\left\langle ij\right\rangle$ denotes nearest neighbors on the single honeycomb layer with $N=2 L\times L$ sites, $\lambda = 1,2$ is the layer index, $\alpha = 1,2,3$ counts the three $3\times 3$ generators $(K^{\alpha})_{\sigma\sigma^{\prime}}=-\rmi \epsilon_{\alpha\sigma\sigma^{\prime}}$ of SO(3), and $\sigma, \sigma' = 1,2,3$ count internal SO(3) degrees of freedom; the diagonal Pauli matrix $\tau^{z}$ acts on the layer degrees of freedom.
The summation over repeated layer and SO(3) indices is implicitly assumed throughout the paper, unless specified otherwise.

The interacting fermion hopping model defined by $\hat{H}$ can be understood as an effective low-energy description of a frustrated Kugel-Khomskii-type spin-orbital model~\cite{seifert20}. In the spin-orbital formulation, the fermionic quasiparticles created by $\hat{c}_{i,\sigma,\lambda}^\dagger$ represent spinons arising from fractionalization in a quantum spin-orbital liquid phase~\cite{chulliparambil20}.
Moreover, in this formulation, the SO(3) vector $\hat{\vec{s}}_{i,\lambda} = (\hat{s}^\alpha_{i,\lambda})$ with components $\hat{s}^\alpha_{i,\lambda} \coloneqq \sum_{\sigma,\sigma'} \hat{c}_{i,\sigma,\lambda}^\dagger K_{\sigma\sigma'}^\alpha \hat{c}_{i,\sigma',\lambda}$ (no summation over $\lambda$) describes the spin density on site $i$ and layer $\lambda$. The interaction term $\hat{H}_{J} = -J \sum_{i} (\hat{\vec{s}}_{i,1} - \hat{\vec{s}}_{i,2})^2$ then favors differences in spin densities between the two layers, and can therefore be understood as a type of interlayer spin-density interaction.
For simplicity, in what follows, we consistently refer to $\hat{H}_{J}$ as spin-density interaction, also in those cases in which no explicit reference to the model's spin-orbital formulation is made.
A sketch of the model is depicted in Fig.~\ref{fig:model}(a).

The interacting bilayer honeycomb model defined by Eqs.~\eqref{eq:model1}-\eqref{eq:model3} has previously been studied at zero temperature using mean-field theory and projective QMC simulations in Ref.~\citep{liu22}.
The corresponding quantum phase diagram as function of interaction strength $J$, as obtained from Ref.~\citep{liu22}, is schematically shown in the bottom panel of Fig.~\ref{fig:model}(b). 
The model features three different phases at zero temperature.
At weak interaction $J < J_\text{c1}$, a fully symmetric Dirac semimetal (DSM) is realized. It features $3 \times 2 \times 2 = 12$ gapless Dirac cones at the Fermi level, arising from the spin, layer, and valley degrees of freedom.
At intermediate interaction $J_\text{c1} < J < J_\text{c2}$, an SO(3)-spin-symmetry-broken Dirac semimetal (SO3 DSM) is stabilized, in which 8 out of the 12 Dirac cones acquire a mass gap, while the 4 leftover Dirac cones remain gapless. This partial gap opening is a consequence of the zero eigenvalue of the SO(3) generators $K^\alpha_{\sigma\sigma'}$.
At strong interaction $J > J_\text{c2}$, a U(1)-charge-symmetry-broken fully gapped state is stabilized. It can be understood as an insulator characterized by U(1) interlayer coherence~\cite{liu22}; however, we show below that the interlayer-coherent insulator (ILC) is degenerate with an $s$-wave superconducting state (SSC) as a consequence of a PPH symmetry.
The fermion single-particle spectral function from mean-field theory~\cite{liu22} is shown for representative values within the three different phases in the top panel of Fig.~\ref{fig:model}(b), illustrating the partial and full gap opening at intermediate and strong coupling, respectively.

\section{Symmetries}
\label{sec:symmetries}

The model's rich phase diagram originates from the large number of fermion internal degrees of freedom together with the intricate symmetry structure.
In the following, we analyze in detail the symmetries of the hopping and interaction parts of the Hamiltonian defined in Eqs.~\eqref{eq:model2} and \eqref{eq:model3}, respectively.
It will prove convenient to do this using a Majorana fermion representation.

\subsection{Majorana representation}

We introduce two Majorana fermions $\hat\gamma_{i,\sigma,\lambda,1}$ and $\hat\gamma_{i,\sigma,\lambda,2}$ for each complex fermion $\hat c_{i,\sigma,\lambda}$ on the two different sublattices $A$ and $B$ as
\begin{align}
\hat{c}_{i,\sigma,\lambda}=
\begin{cases}
\frac{1}{2}\left(\hat\gamma_{i,\sigma,\lambda,1}-\rmi\hat\gamma_{i,\sigma,\lambda,2}\right), & \text{if }  i\in A,\\
\frac{\rmi}{2}\left(\hat\gamma_{i,\sigma,\lambda,1} -\rmi \hat\gamma_{i,\sigma,\lambda,2} \right), &  \text{if } i\in B.
\end{cases}
\end{align}
The Hermitian Majorana operators obey the anticommutation relation
$\{ \hat\gamma_{i,\sigma,\lambda,l},\hat\gamma_{j,\sigma^{\prime}, \lambda^{\prime},l^{\prime}}\} 
= 2 \delta_{i j}\delta_{\sigma \sigma^{\prime}}\delta_{\lambda \lambda^{\prime}}\delta_{l l^{\prime}}$,
where $l =1,2$ corresponds  to the  Majorana index. 
Introducing a twelve-component spinor $\hat{\vec\gamma}_{i}^{\top}= (\hat\gamma_{i,\sigma,\lambda, l })$,
obtained by combining the spin, layer, and Majorana indices into one super-index, allows us to write the kinetic energy  $\hat{H}_{0}$ as
\begin{align}
\hat{H}_{0} & 
=\frac{\rmi t}{4}\sum_{\left\langle i j\right\rangle }\hat{\vec\gamma}_{i}^{\top}\hat{\vec\gamma}_{j}.
\end{align}
In this form, the hopping part of the Hamiltonian becomes manifestly invariant under $\mathrm{O}(12)$ rotations, under which the Majorana spinor transforms as a vector,
\begin{equation}
\hat{\vec\gamma}_{i}    \mapsto    {\hat{\vec\gamma}_i'} = O \hat{\vec\gamma}_{i}, \text{   with   }    O^{\top} O = \mathds{1}_{12}.
\end{equation}

The on-site interaction term becomes in the Majorana formulation
\begin{align}
& \left(\hat{c}_{i,\sigma,\lambda}^{\dagger}K_{\sigma,\sigma^{\prime}}^{\alpha}\tau_{\lambda,\lambda^{\prime}}^{z}\hat{c}_{i,\sigma^{\prime},\lambda^{\prime}}\right)^{2} 
\nonumber \\ &
=  \frac{1}{4} \left[
(\hat{\gamma}_{i,\sigma,\lambda,1}+\rmi \hat{\gamma}_{i,\sigma,\lambda,2})
K_{\sigma \sigma^{\prime}}^{\alpha}\tau_{\lambda \lambda^{\prime}}^{z}
(\hat{\gamma}_{i,\sigma^{\prime},\lambda^{\prime},1}-\rmi \hat{\gamma}_{i,\sigma^{\prime},\lambda^{\prime},2})
\right]^{2}
\nonumber \\ & 
= \frac{1}{4}
\left(\hat{\gamma}_{i,\sigma,\lambda,l} K_{\sigma\sigma^{\prime}}^{\alpha} \tau_{\lambda\lambda^{\prime}}^{z} \hat{\gamma}_{i,\sigma^{\prime},\lambda^{\prime},l}\right)^{2}    
\nonumber \\ & 
= \frac{1}{4} \left(\hat{\vec\gamma}_{i}^{\top} K^{\alpha} \tau^{z}  \mu^{0}\hat{\vec\gamma}_{i}\right)^{2},
\end{align}
where, in the last step, we have introduced the identity matrix $\mu^{0} \coloneqq \mathds{1}_2$ that acts on the Majorana flavor index. 
We recall that $\tau^{z} $ acts  on the  layer  index  and the SO(3) generators $K^{\alpha}$, $\alpha = 1,2,3$,  act  on the spin  indices. 
In the  above,  the  cross  terms  in the Majorana index  vanish  since  the $K^{\alpha}$ are  
antisymmetric and  $\tau^{z}$ is  symmetric.

All in all, the full Hamiltonian  takes  the  form
\begin{equation}
\hat{H}=\frac{\rmi t}{4}\sum_{\left\langle ij\right\rangle}\hat{\vec\gamma}_{i}^{\top}\hat{\vec\gamma}_{j}
- \frac{J}{4}\sum_{i, \alpha}\left(  \hat{\vec\gamma}_{i}^{\top} K^{\alpha}\tau^{z} \mu^{0} \hat{\vec\gamma}_{i}\right)^{2}.
\end{equation}
The  interaction term  reduces  the  O(12)   global  symmetry of the hopping term  down to a subgroup 
that  satisfies
\begin{equation} 
    O^{\top}K^{\alpha}\tau^{z} \mu^{0}  O  =      R^{\alpha \beta} K^{\beta}\tau^{z} \mu^{0},
\end{equation}   
where  $R$  is  an  SO(3)  matrix   and  the  sum over  repeated spin indices  is  implicitly implied.  
We  can now  systematically   read  off  the global  symmetries  of  our   Hamiltonian. 

\subsection{SO(3) spin rotational symmetry}

Here,
\begin{equation}
    O_\text{SO(3)}  =   \rme^{\rmi \theta \vec{e}\cdot \vec{K} },
\end{equation}
with rotation angle $\theta \in [0,2\pi)$
and rotation axis $\vec e$,
and a  similar   form holds  for  $R$. 
Since $O_\text{SO(3)}^{\top} =  O_\text{SO(3)}^{\dagger} $,  the  complex    fermion operators  transform as
\begin{equation}
    \hat{c}_{i,\sigma,\lambda} \mapsto  (O_{\text{SO(3)}})_{\sigma\sigma'} \hat{c}_{i,\sigma',\lambda}.
\end{equation} 
The  SO(3) order  parameter  reads
\begin{align}
\hat{\vec{S}}_i  = (\hat{S}^\alpha_i) \coloneqq \hat{c}_{i,\sigma,\lambda}^{\dagger}K^{\alpha}_{\sigma\sigma'}\hat{c}_{i,\sigma',\lambda},
\end{align}
which can be understood as total spin density, $\hat{\vec{S}}_i = \hat{\vec{s}}_{i,1} + \hat{\vec{s}}_{i,2}$, where $\hat{\vec{s}}_{i,\lambda}$ denotes the spin density at site $i$ on layer $\lambda$.
The spin density $\hat{\vec{S}}_i$ transforms as a vector under SO(3) rotations.

\subsection{U(1) total-charge symmetry}

Here,
\begin{equation}
    O_{\text{U}(1)_\text{T}}  =   \rme^{\rmi \theta \mu^{y}}  
\end{equation}
and $R= 1$. In the above, $\mu^y$ corresponds to the second Pauli matrix that acts on the Majorana flavor index.
For  $i  \in A$, we  have
\begin{equation}
    \hat{c}_{i,\sigma,\lambda}  = \frac{1}{2} \left( \hat{\gamma}_{i,\sigma,\lambda,1}  - \rmi \hat{\gamma}_{i,\sigma,\lambda,2}\right) 
    \mapsto \rme^{\rmi \theta} \hat{c}_{i,\sigma,\lambda},
\end{equation} 
and  a  similar form holds  for  $i \in B$.
A possible order parameter for U(1) total-charge symmetry breaking is the interlayer $s$-wave pairing operator $\hat\Delta_i \coloneqq \hat{c}^\dagger_{i,\sigma,1}\hat{c}^\dagger_{i,\sigma,2}$.

\subsection{U(1) layer-charge symmetry}

Here,
\begin{equation}
    O_{\text{U}(1)_\text{L}}  =   \rme^{\rmi \theta \tau^z \mu^{y}}  
\end{equation}
and $R= 1$.   This symmetry   reflects the  fact  the  charge is  conserved  separately  on  each layer.   
For  the  fermion operator,    the symmetry transformation reads
\begin{equation}
    \hat{c}_{i,\sigma,\lambda}   \mapsto \left(\rme^{\rmi \theta \tau^{z}}\right)_{\lambda \lambda'}\hat{c}_{i,\sigma,\lambda'}.
\end{equation}
An order parameter for spontaneous U(1) layer-charge symmetry breaking is given by the interlayer coherence operator $\hat{n}^\dagger_{i} \coloneqq \hat{c}_{i,\sigma,1}^\dagger \hat{c}^{}_{i,\sigma,2}$.

\subsection{$\mathds{Z}_\mathbf{2}$ partial particle-hole (PPH) symmetry}

Here,
\begin{equation}
    O_{\text{PPH}}  =   \frac{1 +  \tau^z}{2}  +    \frac{1 -  \tau^z}{2} \mu^{z}
\end{equation}
and $R= 1$.     The  PPH symmetry is a  $\mathds{Z}_2$  symmetry since $O_\text{PPH}^2 = 1$. 
It  acts  solely on the  second  layer  where $\tau^{z}_{2,2} = -1$.
In the  fermion  representation, it  leads  to
\begin{align}
    \hat{c}_{i,\sigma,1} & \mapsto  \hat{c}_{i,\sigma,1}, \qquad \text{for all $i \in A,B$}, \label{eq:pph1}\\
    \hat{c}_{i,\sigma,2} & \mapsto  
    \begin{cases}
    \hat{c}^\dagger_{i,\sigma,2}, & \text{if } i \in A,\\
    -\hat{c}^\dagger_{i,\sigma,2}, & \text{if } i \in B,
    \end{cases} \label{eq:pph2}
\end{align}
While  $\det(O_\text{SO(3)})  = \det(O_{\text{U(1)}_\text{T}}) =  \det(O_{\text{U}(1)_\text{L}} )=1$, one  will  show  that $\det(O_\text{PPH}) = -1 $.
A possible  order parameter that probes PPH symmetry  breaking hence reads
\begin{equation}
   \hat{P}_i \coloneqq  \prod_{\sigma,\lambda,l} \gamma_{i,\sigma,\lambda,l},
\end{equation} 
since under O(12)   transformations $\hat{P}_i \mapsto \det(O) \hat{P}_i $.  In  fermion notation, we find
\begin{equation}
    \hat{P}_i = \prod_{\sigma,\lambda} ( 1 - 2 \hat{c}^\dagger_{i,\sigma,\lambda} \hat{c}_{i,\sigma,\lambda}).
\end{equation} 
(In the above equation, there is no summation over $\sigma, \lambda$).
Under the PPH transformation, the interlayer-coherent insulator thus maps onto an $s$-wave superconductor and vice versa.
These two states are therefore degenerate by symmetry. Long-range order in the U(1) layer-charge symmetry broken sector thus implies long-range order  in the U(1) total-charge sector.
We note that the insertion of a flux quantum, as done in our previous work~\cite{liu22}, breaks the PPH symmetry, and as such lifts on finite lattices the degeneracy between the interlayer-coherent insulator and the $s$-wave superconductor in favor of the insulating state. In this work, we therefore refrain from inserting a flux quantum, such that the PPH-symmetry-required degeneracy remains intact already on finite lattices.

\section{QMC simulations}
\label{sec:QMC}

In this section, we present aspects of our QMC simulations. 
It is beyond the scope of this article to provide a detailed account of the algorithm and we will concentrate on model specific issues. We start by describing two different possible Trotter decompositions, demonstrate the absence of the sign problem, describe the specific implementation, and finally illustrate the convergence properties with respect to the limit of small Trotter time steps of the two different decompositions.

\subsection{Trotter decomposition}

We represent the partition function at inverse temperature $\beta$ as 
\begin{align}
Z & = \Tr \prod_{n=1}^{M}\exp(-\Delta\tau\hat{H}),
\end{align}
where $M$ corresponds to the number of Trotter time steps $\Delta \tau = \beta/M$.

\paragraph{Naive Trotter decomposition.}
For sufficiently large $M \gg 1$, the partition function can be decomposed as
\begin{align}
\exp(-\Delta\tau\hat{H}) & =
\int \mathcal D\vec\phi  \,\rme^{-\sum_i \frac{ \vec\phi_{i,\tau}^2}{2}}
\Tr \prod_{n=1}^M \Biggl\{ \rme^{-\frac{\Delta\tau}{2}\hat{H}_{0}} \times
\nonumber \\ & \  
\left(\prod_{\alpha=1}^{3}\rme^{-\sum_{i,j}\ve{\hat{c}}_{i,\lambda}^{\dagger}V^{\alpha}_{ij}(\vec\phi) {\tau^z_{\lambda\lambda'}} \ve{\hat{c}}_{j,\lambda'}}\right)
\rme^{-\frac{\Delta\tau}{2}\hat{H}_{0}} \Biggr\}
+ \mathcal{O}(\Delta\tau^{2}),
\label{eq:naive-trotter}
\end{align}
where the functional integral $\int \mathcal D\vec\phi$ is assumed over the auxiliary fields $\vec\phi_{i,\tau} = (\phi^\alpha_{i,\tau})$, and we have introduced $V^\alpha_{ij}(\vec\phi) = \sqrt{2\Delta \tau J}  \delta_{ij}  \phi_{i,\tau}^\alpha K^\alpha$ and $\mathbf{\hat{c}}_{i,\lambda} = (\hat{c}_{i,\sigma,\lambda})$.
The partition function can then be represented in terms of a fermion determinant as
\begin{align}
Z & =  \int \mathcal D\vec\phi \rme^{-\sum_{i}\frac{\vec\phi_{i,\tau}^{2}}{2}} \det W(\vec\phi)
\end{align}
with the fermion matrix given as
\begin{align}\label{eq:fermion-matrix-naive}
W(\vec \phi) = \mathds{1} + \prod_{n=1}^{M} \left\{
\rme^{-{\frac{\Delta \tau}{2}T}} \left(\prod_{\alpha=1}^{3}\rme^{-V^{\alpha}(\vec\phi)}\right) \rme^{-{\frac{\Delta\tau}{2}T}} \right\}.
\end{align}
Here, $V^\alpha = (V_{ij}^\alpha) \tau^z$ corresponds to the vertex matrix and $T = (T_{ij}) \tau^0$ corresponds to the hopping matrix, with elements $T_{ij} = -t$ if $i$ and $j$ are nearest neighbors and $T_{ij} = 0$ otherwise, and the identity matrix $\tau^0 = \mathds{1}_2$ acts on the layer degrees of freedom.
As the generators of SO(3), $K^\alpha$, $\alpha = 1, 2, 3$, do not commute with each other, the naive Trotter decomposition in Eq.~\eqref{eq:naive-trotter} leads to a Hermitian time evolution only in the limit $M \to \infty$.

\paragraph{Hermitian Trotter decomposition.}
A Hermitian time evolution already at finite $M$ can be achieved at the expense of introducing another set of auxiliary fields $\vec{\chi}$ in a symmetric Trotter decomposition as
\begin{align}
\exp(-\Delta\tau\hat{H}) & =
\int \mathcal D\vec\phi \mathcal D\vec{\chi}  \,\rme^{-\sum_i \frac{ \vec\phi_{i,\tau}^2 +  \vec\chi_{i,\tau}^2}{2}}
\Tr \prod_{n=1}^M \Biggl\{ \rme^{-\frac{\Delta\tau}{2}\hat{H}_{0}} \times
\nonumber \\ & \quad 
\left(\prod_{\alpha=1}^{3}\rme^{-\sum_{i,j}\frac12 \ve{\hat{c}}_{i}^{\dagger}V^{\alpha}_{ij} (\vec\phi) \ve{\hat{c}}_{j}}\right) \times
\nonumber \\ & \quad 
\left(\prod_{\beta=3}^{1}\rme^{-\sum_{i,j}\frac12 \ve{\hat{c}}_{i}^{\dagger} V^{\beta}_{ij} (\vec{\chi}) \ve{\hat{c}}_{j}}\right)
\rme^{-\frac{\Delta\tau}{2}\hat{H}_{0}} \Biggr\}
\nonumber \\ & \quad 
+ \mathcal{O}(\Delta\tau^{2}).
\end{align}
The partition function can then analogously be represented in terms of a fermion determinant $\det W(\vec\phi,\vec\chi)$, with the corresponding fermion matrix given as
\begin{align}\label{eq:fermion-matrix-symmetric}
W(\vec \phi, \vec{\chi}) & = \mathds{1} + \prod_{n=1}^{M} \Biggl\{
\rme^{-\frac{\Delta \tau T}{2} }
\left(\prod_{\alpha=1}^{3}\rme^{-\frac{V^{\alpha}(\vec\phi)}{2}} \right)
 \left(\prod_{\beta=3}^{1}\rme^{-\frac{V^{\beta}(\vec\chi)}{2}}\right) \times
\nonumber \\ & \quad 
\rme^{\frac{-\Delta \tau T}{2} } \Biggr\}.
\end{align}
Below, we show that the above Hermitian Trotter decomposition leads to significantly improved convergence properties towards the $M \to \infty$ limit.

\subsection{Absence of sign problem}

For simplicity, we show the absence of the sign problem for the naive Trotter decomposition with the fermion determinant $\det W(\vec\phi)$. Analogous arguments hold for the Hermitian Trotter decomposition with the fermion determinant $\det W(\vec\phi,\vec\chi)$.

In our model, the hopping matrix $T$ and the vertex matrix $V^\alpha(\vec\phi)$ are block diagonal with respect to the layer index, such that the fermion determinant can be written as a product of single-layer fermion determinants as
\begin{align}
\det W(\vec\phi) = \det W_{1}(\vec\phi) \det W_{2}(\vec \phi),
\end{align}
where $W_\lambda$ corresponds to the fermion matrix on the $\lambda$-th layer, $\lambda = 1,2$,
\begin{align}
W_\lambda(\vec\phi) = \mathds{1} + \prod_{n=1}^{M} \left\{
\rme^{\frac{-\Delta \tau(T_{ij})}{2}} \left(\prod_{\alpha=1}^{3}\rme^{(-1)^\lambda(V_{ij}^{\alpha})(\vec\phi)}\right) \rme^{\frac{ -\Delta \tau (T_{ij})}{2}} 
\right\},
\end{align}
where the $N \times N$ matrices $(T_{ij})$ and $(V_{ij}^{\alpha})$ act only on a single layer. As the hopping matrix is real, $(T_{ij})^* = (T_{ij})$, and the vertex matrix purely imaginary, $(V^{\alpha}_{ij})^* = -(V^{\alpha}_{ij})$, we have $[\det W_1(\vec\phi)]^* = \det W_2(\vec\phi)$ for all real configurations $\vec\phi$.
As a consequence, the fermion determinant is nonnegative,
\begin{align}
\det W(\vec\phi) \geq 0,
\end{align}
facilitating sign-problem-free QMC simulations.

\subsection{Specific implementation}

We use the ALF  implementation \cite{ALF_v2} of   the  auxiliary field  QMC algorithm \cite{blankenbecler81,white89,sugiyama86,sorella89}.   
This package utilizes Gauss-Hermite quadrature to replace sampling over continuous fields with a discrete field that takes the values $\pm 2$ and $\pm 1$. For further details, we refer to Refs.~\cite{ALF_v2,Assaad97}. While both approaches are formally equivalent, sampling over discrete fields generally reduces fluctuations.
Our finite inverse  temperature $\beta$ calculations  are  carried out  in the grand-canonical ensemble, choosing linear system sizes $L=6,9,12,15,18$ with periodic boundary conditions. 
For the zero-temperature results, we adopt a $\beta = L$ scaling, consistent with a dynamical critical exponent $z=1$, as naively expected from the linear fermion dispersion.
If not stated otherwise, we use a Trotter time step of $\Delta \tau = 0.2$.
We choose units in which $t=1$ and $k_\mathrm{B} = 1$, such that the temperature $T = 1/\beta$ and the coupling constant $J\geq 0$ become dimensionless parameters, which we scan.

\subsection{Convergence of Trotter decomposition}

\begin{figure}[tb]
\includegraphics[width=\linewidth]{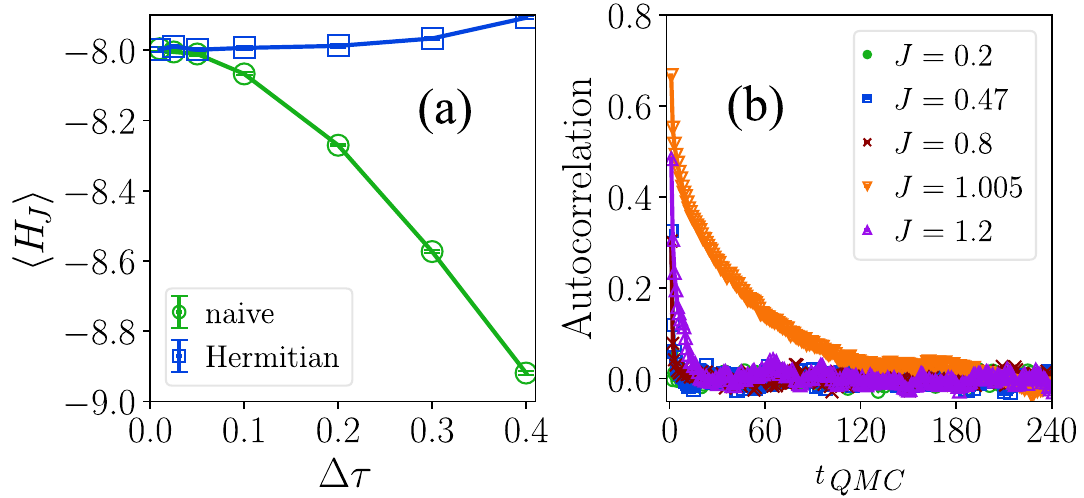}
\caption{%
(a)~Expectation value of the interaction part of the Hamiltonian $\langle \hat H_{J}\rangle$ as function of Trotter time step $\Delta\tau$ for the two different Trotter decompositions, using $t=0$, $J=1$, and $\beta=4$. The Hermitian Trotter decomposition leads to significantly improved convergence properties towards the limit $\Delta \tau = \beta/M \to 0$.
(b)~Autocorrelation function of hopping part of the Hamiltonian $\langle \hat{H}_{0}\rangle$ as function of Monte Carlo time $t_{\text{QMC}}$ for different fixed values of $J$, using $L=6$ and $\beta = 18$. The autocorrelation time remains small near the GN-SO(3) transition point at $J_\text{c1} = 0.465(2)$, but significantly increases in the vicinity of the SO(3)-U(1) transition point at $J_\text{c2} = 1.057(10)$.
\label{fig:QMC-auto-test}}
\end{figure}

Figure~\ref{fig:QMC-auto-test}(a) shows the expectation value of the interaction part of the Hamiltonian $\langle \hat H_{J} \rangle$ as function of the Trotter time step $\Delta\tau$ for the two different Trotter decompositions.
Importantly, the Hermitian decomposition leads, in comparison with the naive decomposition, to significantly improved convergence properties towards the limit $\Delta \tau = \beta/M \to 0$. In the remainder of this paper, we therefore exclusively employ the Hermitian Trotter decomposition. As shown below, this allows us to obtain significantly improved estimates for the quantum critical properties at zero temperature in comparison with our previous work~\cite{liu22}.

\section{Zero-temperature results}
\label{sec:zerotemperature}

In this section, we demonstrate that the above-described advances in the implementation of our model leads, in comparison with our previous work~\cite{liu22}, to significant improvements in the results obtained at zero temperature.
This is in particular true in the vicinity of the two quantum phase transition points.
%

\subsection{GN-SO(3) transition}

We start by discussing the GN-SO(3) transition at $J_\text{c1}$ between the fully symmetric Dirac semimetal and SO(3)-spin-symmetry-broken partially gapped Dirac semimetal. 
To this end, we measure the SO(3) order parameter $m_\text{SO(3)}$ defined via 
\begin{equation}
	m_{\text{SO(3)}}^{2}(J,L)=\frac{\mathcal S_{\text{SO(3)}}(\boldsymbol{k}=\boldsymbol\Gamma,\tau=0)}{L^{2}},
\end{equation}
where 
\begin{align}
\mathcal S_{\text{SO(3)}}(\boldsymbol{k},\tau) = \sum_{i} \rme^{-\rmi \boldsymbol{k} \cdot \boldsymbol{r}_i} \left\langle \hat{\vec{s}}_{i,\lambda}(\tau) \cdot \hat{\vec{s}}_{0,\lambda}(0) \right\rangle
\end{align}
is the SO(3) spin structure factor at momentum $\boldsymbol{k}$ and imaginary time $\tau$, and $\boldsymbol{r}_i$ denotes the position vector of the lattice site $i$.
The spin structure factor describes correlations between the spin density $\hat{\vec{s}}_{i,\lambda}(\tau) = \mathbf{\hat{c}}^\dagger_{i,\lambda}(\tau) \vec K \mathbf{\hat c}^{\phantom\dagger}_{i,\lambda}(\tau)$ (no summation over $\lambda$), where $\mathbf{\hat{c}}^\dagger_{i,\lambda}(\tau)$ is the time-evolved fermion operator in the Heisenberg picture.
In order to locate the GN-SO(3) transition, we compute the renormalization group invariant SO(3) correlation ratio~\cite{kaul15}
\begin{align}
R_\mathrm{c}^{\text{SO(3)}}(J,L)=1-\frac{\mathcal S_{\text{SO(3)}}(\boldsymbol{k}=\boldsymbol\Gamma+\rmd\boldsymbol{k},\tau=0)}{\mathcal S_{\text{SO(3)}}(\boldsymbol{k}=\boldsymbol{\Gamma},\tau=0)},
\end{align}
where $\rmd \boldsymbol k$ connects neighboring momenta in the Brillouin zone of the finite-size lattice.
In the limit of large system size, the correlation ratio $R_\mathrm{c}^\text{SO(3)}$ goes to zero (one) in the SO(3)-ordered (disordered) phase. 
The crossings of the correlation-ratio curves as function of the tuning parameter $J$ for different fixed lattices sizes $L$ indicate the location of the SO(3)-spin-symmetry-breaking phase transition.
In addition, we measure the fermion quasiparticle weight 
\begin{align}
Z_\text{qp}(J,L)=\frac{G(J,L)}{G(0,L)}
\end{align}
where $G(J,L)$ is constructed from the time dependence of the fermion Green's function as
\begin{align}
G(J,L)=\left\langle \mathbf{\hat{c}}_{{i},\lambda}^{\dagger}({\beta}/{2}) \cdot \mathbf{\hat{c}}_{{i},\lambda}(0) \right\rangle.
\end{align}

\begin{figure}[tb]
\begin{centering}
\includegraphics[width=\columnwidth]{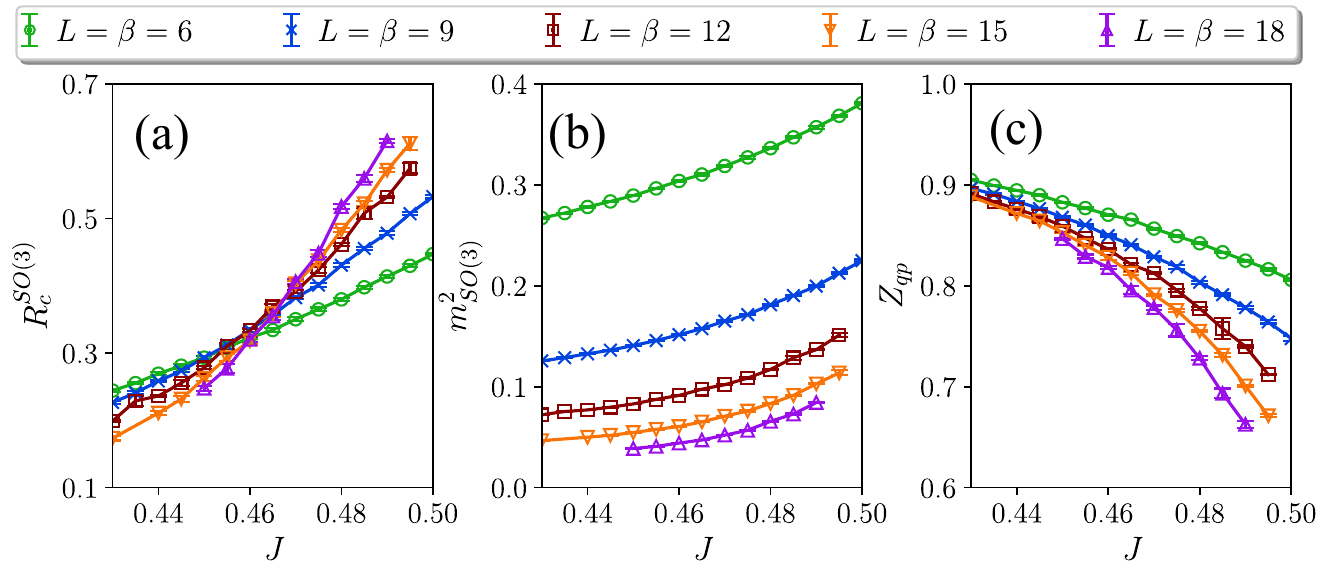}
\par\end{centering}
\caption{%
(a)~SO(3) correlation ratio $R_\mathrm{c}^{\text{SO(3)}}$ as function of coupling $J$ in the vicinity of the GN-SO(3) transition for different fixed lattice sizes $L$. The crossing point of the different finite-size curves indicates the location of the GN-SO(3) quantum critical point. 
(b)~Same as (a), but for the SO(3) order parameter $m_{\text{SO(3)}}^2$.
(c)~Same as (a), but for the fermion quasiparticle weight $Z_\text{qp}$.
\label{fig:GN-SO3-rawdata}}
\end{figure}

Figure~\ref{fig:GN-SO3-rawdata} shows the raw data of these three observables near the GN-SO(3) quantum phase transition. 
For the correlation ratio [Fig.~\ref{fig:GN-SO3-rawdata}(a)], we observe a clear crossing point, indicating the position of the GN-SO(3) quantum critical point. 
Following the finite-size scaling hypothesis~\citep{campostrini14}, the observables are expected to obey the critical scaling form
\begin{align}
R_{\mathrm{c}}^{\mathrm{SO(3)}}(J,L) & \sim f_{0}^{R}(jL^{1/\nu})+L^{-\omega}f_{1}^{R}(jL^{1/\nu}), \\
m_{\text{SO(3)}}^{2}(J,L) & \sim L^{-1-\eta_{\phi}}[f^{m}_0(jL^{1/\nu})+L^{-\omega}f_{1}^{m}(jL^{1/\nu})], \\
Z_\text{qp}(J,L) & \sim L^{-\eta_{\psi}}[f^{z}_0(jL^{1/\nu})+L^{-\omega}f^{z}_1(jL^{1/\nu})],
\end{align}
where $j = J - J_\mathrm{c1}$ corresponds to the reduced coupling and $\omega$ denotes the correction-to-scaling exponent.
We use two different types of finite-size analysis in order to extract the critical coupling and the corresponding exponents. The two methods lead to results that are consistent within the numerical uncertainty.

\begin{figure}[tb]
\includegraphics[width=\columnwidth]{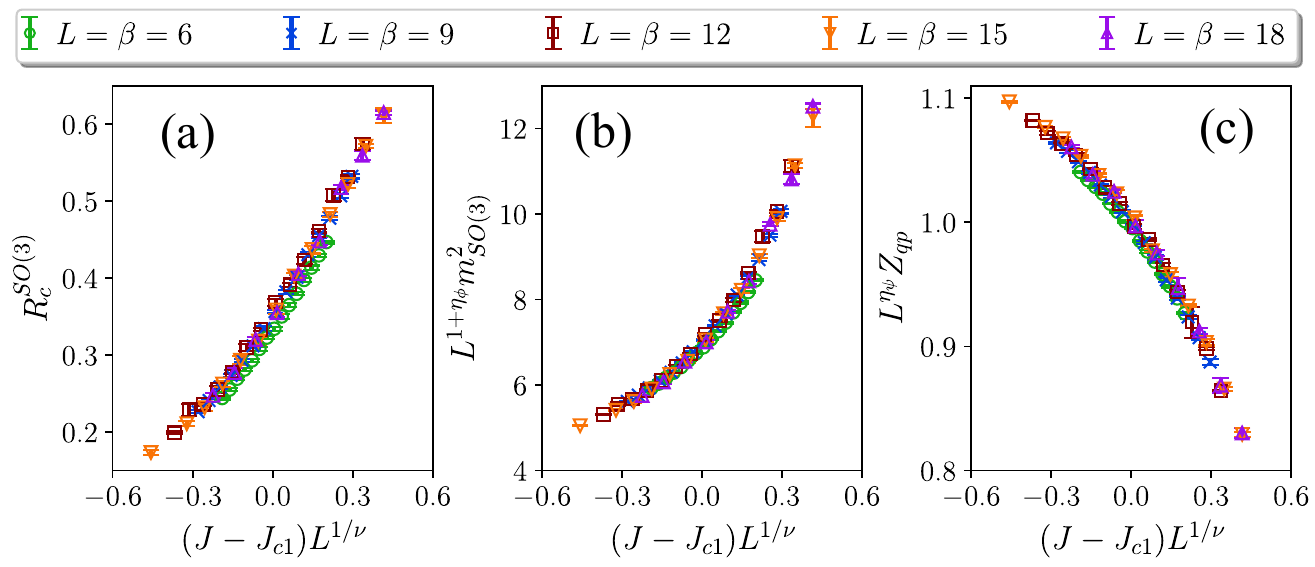}
\caption{%
(a)~Finite-size scaling collapse of SO(3) correlation ratio as function of $(J-J_\mathrm{c1}) L^{1/\nu}$ in the vicinity of the GN-SO(3) transition, yielding the best-fit parameters $J_\text{c1}=0.465(2)$ and $1/\nu=0.86(8)$.
(b)~Same as (a), but for the order parameter $m_{\text{SO(3)}}^{2}$ in units of $L^{-1-\eta_\phi}$, yielding $\eta_{\phi}=0.73(2)$.
(c)~Same as (a), but for the fermion quasiparticle weight $Z_\text{qp}$ in units of $L^{-\eta_\psi}$, yielding $\eta_{\psi}=0.078(8)$.
\label{fig:GN-SO3-collapse-data}}
\end{figure}

\paragraph{Data-collapse analysis.}
As a first step, we ignore the corrections to scaling $\propto \mathcal O(L^{-\omega})$ in the above finite-size scaling forms.
In order to extract the critical coupling $J_\text{c1}$ and the correlation-length exponent $\nu$, we fit the correlation ratio $R_{\mathrm{c}}^{\mathrm{SO(3)}}(J,L)$ as function of $(J-J_\mathrm{c1}) L^{1/\nu}$ to a fourth-order polynomial. 
The optimal scaling collapse is obtained for the best-fit parameters $J_\text{c1}=0.465(2)$ and $1/\nu=0.86(8)$, see Fig.~\ref{fig:GN-SO3-collapse-data}(a).
From the finite-size scalings of the order parameter $m_{\text{SO(3)}}^{2}(j,L)$ and the $Z_\text{qp}(j,L)$, we analogously obtain the estimates $\eta_{\phi}=0.73(2)$ for the order-parameter anomalous dimension and $\eta_{\psi}=0.078(8)$ for the fermion anomalous dimension, respectively. The corresponding scaling collapses are shown in Figs.~\ref{fig:GN-SO3-collapse-data}(b) and \ref{fig:GN-SO3-collapse-data}(c).

\begin{figure}[tb]
\begin{centering}
\includegraphics[width=\columnwidth]{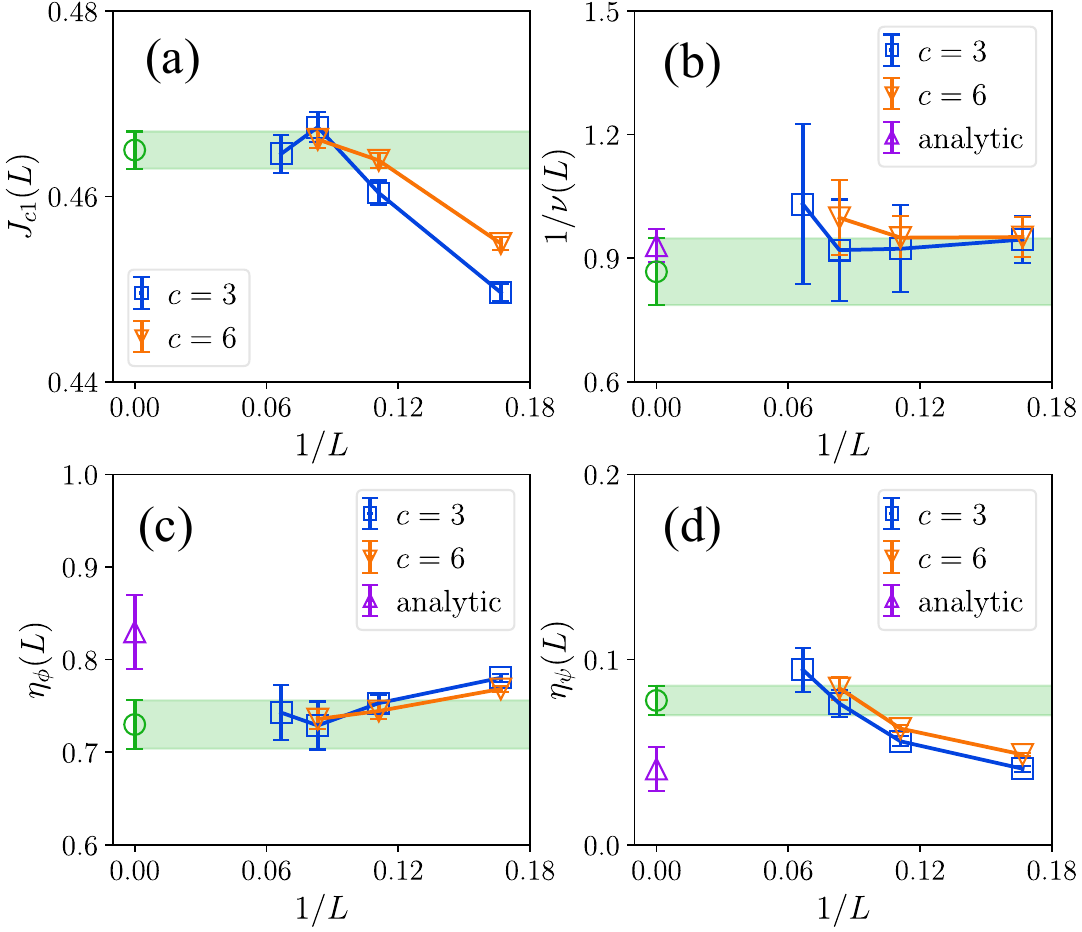}
\par\end{centering}
\caption{%
(a)~Finite-size critical coupling $J_\text{c1}(L)$ as function of $1/L$ from crossing-point analysis using size increments $c = 3$ (blue squares) and $c = 6$ (orange diamonds), respectively, in comparison with the critical coupling $J_\text{c1}$ obtained from the data-collapse analysis (green dot). For sufficiently large lattice sizes, the estimates are consistent with each other within the numerical uncertainty.
(b)~Same as~(a), but for the effective correlation-length exponent $1/\nu(L)$.
(c)~Same as~(a), but for the effective order-parameter anomalous dimension $\eta_\phi(L)$.
(d)~Same as~(a), but for the effective fermion anomalous dimension $\eta_\psi(L)$.
Purple triangles in~(b)--(d) show the field-theoretical estimates from Ref.~\cite{ray21} for comparison.
\label{fig:GNSO3-crossing-point}}
\end{figure}

\paragraph{Crossing-point analysis.}
As a second step, we compare the above results of the scaling-collapse analysis with those of a crossing-point analysis that takes scaling-correction effects into account~\cite{shao16}.
We define the finite-size critical coupling $J_\text{c1}(L)$ as the crossing point of the SO(3) correlation ratio $R_\text{c}^{\text{SO(3)}}(J,L)$ of system sizes $L$ and $L+c$ with size increment $c$ as
\begin{align}
R_\text{c}^{\text{SO(3)}}(J_\text{c1}(L),L)=R_\text{c}^{\text{SO(3)}}(J_\text{c1}(L),L+c).
\end{align}
For increasing system sizes, the finite-size critical coupling approaches the thermodynamic critical point as 
$J_\text{c1}(L)=J_\text{c1}+ a L^{-\omega - 1/\nu}$,
with nonuniversal coefficient $a$~\cite{toldin15}.
Figure~\ref{fig:GNSO3-crossing-point}(a) shows the finite-size critical coupling as function of $1/L$ for two different size increments (orange diamonds and blue squares) in comparison with the critical coupling obtained from the data-collapse analysis (green circle).
For sufficiently large lattice sizes, the estimates are consistent with each other within the numerical uncertainty.

Having computed the finite-size critical coupling, effective finite-size critical exponents can be obtained from
\begin{align}
1/\nu(L) & ={\ln \frac{s(J_\text{c1}(L),L+c)}{s(J_\text{c1}(L),L)}} \left/\, {\ln\frac{L+c}{L}} \right., \label{eq:crossing-point-nu}\\
\eta_{\phi}(L) & = - 1 - {\ln \frac{m^{2}_\text{SO(3)} (J_\text{c1}(L),L+c)}{m^{2}_\text{SO(3)}(J_\text{c}(L),L)}} \left/\, {\ln\frac{L+c}{L}} \right.,\\
\eta_{\psi}(L) & =-{\ln \frac{Z_\text{qp}(J_\text{c1}(L),L+c)}{Z_\text{qp}(J_\text{c1}(L),L)}} \left/\, {\ln\frac{L+c}{L}} \right.,
\end{align}
where $s(J,L)= \partial R_{\mathrm{c}}^{\mathrm{SO(3)}}(J,L) / \partial J$
corresponds to the slope of the correlation ratio as function of the coupling $J$.
For increasing system sizes $L$, the deviation between the effective critical exponent and the corresponding value in the thermodynamic limit vanishes with $L^{-\omega}$.
Our results for the effective finite-size critical exponents as function of $1/L$ for two different size increments (blue squares and orange diamonds) in comparison with the critical exponents obtained from the data-collapse analysis (green circles) are shown in Figs.~\ref{fig:GNSO3-crossing-point}(b)--(d).
For sufficiently large lattice sizes, the estimates from the different analyses are again consistent with each other within the numerical uncertainty.

From the symmetry of the order parameter and the low-energy field content of the lattice model, we expect the continuous quantum phase transition at $J_\mathrm{c1}$ to be described by the GN-SO(3) field theory~\cite{seifert20} given by the action $S = \int d^3 x \mathcal L$ with
\begin{align}
\mathcal L = \bar\Psi \gamma^\mu\partial_\mu \Psi - g \vec\phi \cdot \bar\Psi (\mathds{1}_4 \otimes \vec{K}) \Psi,
\end{align}
where $\vec \phi = (\phi^\alpha)$, $\alpha = 1,2,3$, corresponds to the SO(3) vector order parameter and the complex fermion fields $\Psi$ and $\bar\Psi$ have $2^3 \times 3 = 24$ components, arising from sublattice, valley, layer, and internal SO(3) degrees of freedom. The Dirac matrices $\gamma^\mu$, $\mu = 0,1,2$, form a 24-dimensional representation of the Clifford algebra, and the summation convention over repeated space-time indices $\mu$ is assumed.
The GN-SO(3) field theory has previously been studied using three-loop $4-\epsilon$ expansion, next-to-leading-order $1/N$ expansion, and functional renormalization group calculations in the local-potential approximation~\cite{ray21}. The corresponding estimates for the critical exponents from combining these analytical methods are also shown in Figs.~\ref{fig:GNSO3-crossing-point}(b)--(d) for comparison (purple triangles).
Importantly, in comparison with our previous work~\cite{liu22}, in which the naive Trotter decomposition has been employed, the improved implementation using the Hermitian Trotter decomposition leads to results that are significantly closer to the field-theoretical estimates, in particular for the anomalous dimensions $\eta_\phi$ and $\eta_\psi$.
We attribute the remaining small deviations of the order of $\lesssim 2\sigma$ to systematic uncertainties that are difficult to control within both field-theoretical and numerical approaches.

\subsection{SO(3)-U(1) transition}

We continue by describing the behavior of the system near the potentially deconfined metallic transition at $J_\text{c2}$ between the SO(3)-spin- and U(1)-charge-symmetry-broken states.
Across this transition, the remaining gapless fermion modes acquire a spectral gap.
In order to characterize the transition, we compute both the SO(3) spin structure factor $\mathcal S_\text{SO(3)}(\boldsymbol{k},\tau)$ and the U(1) layer-charge structure factor, defined as
\begin{align} \label{eq:U(1)-structure-factor}
\mathcal S_{\mathrm{U}(1)}(\boldsymbol k, \tau) = \sum_{{i}}\rme^{-\rmi \boldsymbol{k} \cdot \boldsymbol{r}_{i}} 
\left\langle 
[ \hat n_i^\dagger(\tau) + \hat n_i^{\phantom\dagger}(\tau) ]
[ \hat n_0^\dagger(0) + \hat n_0^{\phantom\dagger}(0) ]
\right\rangle ,
\end{align}
where $n_i^\dagger(\tau) = \hat{c}_{i,\sigma,1}^\dagger(\tau) \hat{c}^{\phantom\dagger}_{i,\sigma,2}(\tau)$ corresponds to the interlayer coherence operator in the Heisenberg picture.
Under the PPH transformation, the U(1) layer-charge structure factor transforms into the U(1) pairing structure factor 
\begin{align}
\mathcal S'_\mathrm{U(1)}(\boldsymbol k, \tau) = \sum_{{i}}\rme^{-\rmi \boldsymbol{k} \cdot \boldsymbol{r}_{i}} 
\left\langle 
[ \hat \Delta_i^\dagger(\tau) + \hat \Delta_i^{\phantom\dagger}(\tau) ]
[ \hat \Delta_0^\dagger(0) + \hat \Delta_0^{\phantom\dagger}(0) ]
\right\rangle ,
\end{align}
where $\Delta_i(\tau) = \hat{c}^\dagger_{i,\sigma,1}(\tau) \hat{c}^\dagger_{i,\sigma,2}(\tau)$ corresponds to the interlayer $s$-wave pairing operator in the Heisenberg picture.
In this work, we use a microscopic implementation that explicitly preserves PPH symmetry. As a consequence,  our results presented for $\mathcal S_{\mathrm{U}(1)}(\boldsymbol k, \tau)$ in the following are representative for both the U(1) layer-charge and U(1) pairing structure factors.
The U(1) correlation ratio is defined as
\begin{align}
R_\mathrm{c}^{\text{U(1)}}(J,L)=1-\frac{\mathcal S_{\text{U(1)}}(\boldsymbol{k}=\boldsymbol\Gamma+\rmd\boldsymbol{k},\tau=0)}{\mathcal S_{\text{U(1)}}(\boldsymbol{k}=\boldsymbol{\Gamma},\tau=0)},
\end{align}
Analogously, we define the PPH structure factor $\mathcal S_\text{PPH}(\boldsymbol k, \tau)$ and the corresponding correlation ratio $R_\mathrm{c}^\text{PPH}$ using the parity operator $\hat P_i(\tau) = \prod_{\sigma,\lambda} [1 - 2 \hat{c}^\dagger_{i,\sigma,\lambda}(\tau) \hat{c}^{}_{i,\sigma,\lambda}(\tau)]$ (no summation over $\sigma, \lambda$).

\begin{figure}[tb]
\begin{centering}
\includegraphics[width=\columnwidth]{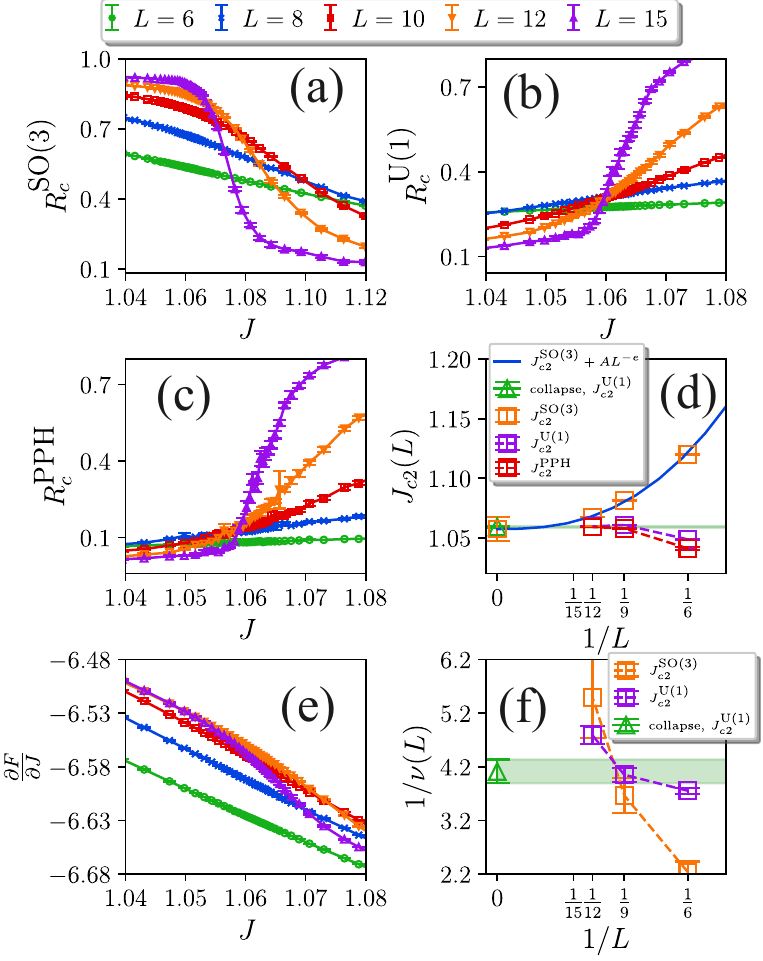}
\par\end{centering}
\caption{%
(a)~SO(3) correlation ratio $R_\mathrm{c}^\text{SO(3)}$ as function of $J$ in the vicinity of the SO(3)-U(1) transition for different fixed lattice sizes $L$.
(b)~Same as (a), but for the U(1) correlation ratio $R_\mathrm{c}^\text{U(1)}$.
(c)~Same as (a), but for the PPH correlation ratio $R_\mathrm{c}^\text{PPH}$.
(d)~Finite-size critical coupling $J_\mathrm{c2}(L)$ as function of $1/L$ from crossing-point analysis of SO(3) correlation ratio (orange squares), U(1) correlation ratio (purple squares), and PPH correlation ratio (red squares).
Blue solid curve indicates a power-law fit of $J_\mathrm{c2}(L) = J_\mathrm{c2} + a L^{-e}$ from SO(3) correlation ratio, yielding $J_\mathrm{c2}=1.057(10)$.
Critical coupling $J_\mathrm{c2}$ from data-collapse analysis of U(1) correlation ratio is also shown for comparison (green triangle).
All values for $J_\mathrm{c2}$ are consistent with each other, indicating a direct SO(3)-U(1) transition without an intermediate coexistence phase.
(e)~First derivative of free energy $F$ as function of $J$, showing no discontinuity near $J_\mathrm{c2}$ within our accuracy.
(f)~Effective finite-size critical exponent $1/\nu(L)$ as function of $1/L$ from crossing-point analysis of SO(3) correlation ratio (orange squares) and U(1) correlation ratio (purple squares), in comparison with estimate for $1/\nu$ from data-collapse analysis of U(1) correlation ratio (green triangle).
\label{fig:dqcp-ana}}
\end{figure}

Figures~\ref{fig:dqcp-ana}(a)--(c) show our results for the SO(3), U(1) and PPH correlation ratios. These observables indicate a phase transition between the partially-gapped SO(3)-spin-symmetry-broken semimetal at $J_\mathrm{c1} < J < J_\mathrm{c2}$ and the interlayer-coherent insulating or $s$-wave superconducting state, which breaks U(1)-charge and PPH symmetries, at $J > J_\mathrm{c2}$.
This interpretation is also supported by the finite-size scaling of the correlation-ratio crossing points for consecutive system sizes, shown as function of $1/L$ in Fig.~\ref{fig:dqcp-ana}(d). Importantly, the finite-size critical couplings $J_\text{c2}(L)$ associated with the three different order parameters all scale towards a unique limiting value of $J_\text{c2} = 1.057(10)$ (orange triangle at $1/L = 0$). 
This indicates a direct transition at $J_\mathrm{c2}$ between the two different symmetry-broken states, without an intermediate coexistence phase.
In Fig.~\ref{fig:dqcp-ana}(d), we present the first derivative
of the free energy $F$ as function of $J$ in the vicinity of $J_{c2}$. Up to the largest system
sizes considered ($L \leq 15$), no discontinuity in $\partial F/\partial J$ can be identified, in sharp contrast to the mean-field result~\cite{liu22}. This suggests a fluctuation-induced continuous or weakly-first-order transition at $J_\mathrm{c2}$. 
In order to characterize the associated quantum critical behavior, we attempt a scaling collapse of $R_\mathrm{c}^\text{U(1)}(J,L)$ as function of $(J-J_\mathrm{c2}) L^{1/\nu}$. The resulting values of $J_\mathrm{c2}$ and $1/\nu$ are shown in Figs.~\ref{fig:dqcp-ana}(d) and \ref{fig:dqcp-ana}(f), respectively (green triangles).
We note that the value of $J_\mathrm{c2}$ from the data-collapse analysis is consistent with the extrapolation of the correlation-ratio crossing points [Fig.~\ref{fig:dqcp-ana}(d)].
For comparison, Fig.~\ref{fig:dqcp-ana}(f) also shows estimates for the correlation-length exponent $1/\nu$ from a crossing-point analysis analogous to Eq.~\eqref{eq:crossing-point-nu}, replacing $J_\mathrm{c1} \to J_\mathrm{c2}$ therein and choosing $c = 3$.
Several remarks are in order:
First, the autocorrelation time in the QMC simulations significantly increases near $J_\mathrm{c2}$, as illustrated in Fig.~\ref{fig:QMC-auto-test}(b). This restricts the available system sizes in the vicinity of $J_\mathrm{c2}$ to $L \leq 15$.
Second, for the available system sizes, there is a significant drift in the estimates for $1/\nu$, suggesting the presence of sizable corrections to scaling. 
Third, we note that the finite-size values for $1/\nu(L)$ from the SO(3) and U(1) correlation ratios are consistent with each other within the numerical uncertainty, pointing towards a unique value for $1/\nu$ from the two different correlation functions.
Fourth, if we assume a dynamical exponent $z=1$, the estimates for $1/\nu$ are above the value $d+z$ expected for a first-order transition~\cite{nienhuis75, fisher82, binder87, mueller14, francini24}.
These results are again consistent with our interpretation of a fluctuation-induced continuous or weakly-first-order order-to-order transition at $J_\mathrm{c2}$.
We refrain from estimating fermion and order-parameter anomalous dimensions, as the corresponding fitting processes in the scaling-collapse analysis involve an additional free parameter and are therefore even harder to control.

\begin{figure}[tb]
\begin{centering}
\includegraphics[width=\columnwidth]{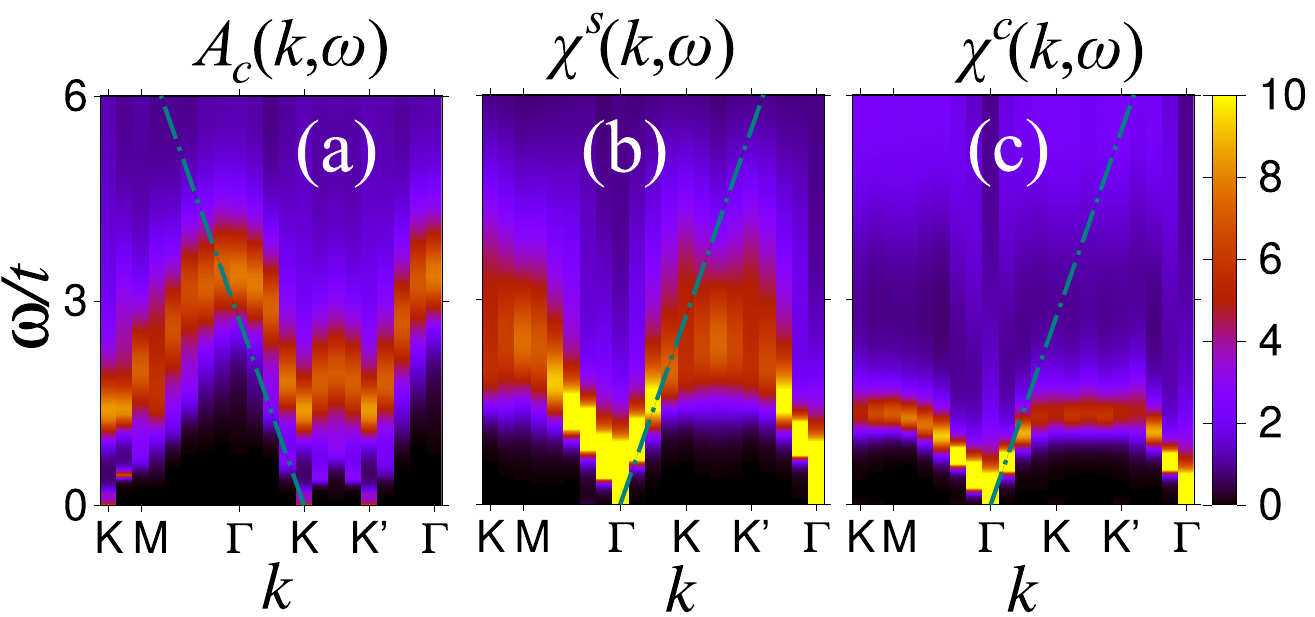}
\par\end{centering}
\caption{%
Spectral functions in
(a)~single-particle, 
(b)~particle-hole,
and
(c)~particle-particle
channels at the SO(3)-U(1) transition point for $J = 1.057 \approx J_\text{c2}$.
Here, we show data for system size $L=12$ and inverse temperature $\beta = 24$.
Dash-dotted lines indicate the linear dispersions of the gapless excitations around the $\mathbf K$ (single-particle channel) and $\boldsymbol \Gamma$ (particle-hole and particle-particle channels) points in the Brillouin zone. The data support emergent Lorentz symmetry characterized by a single ``velocity of light.''}
\label{fig:spectral-functions}
\end{figure}

Spectral functions in different  channels and  at the  critical point provide valuable insight  on  emergent  symmetries.  For  instance,  Lorentz  symmetry  imposes the  constraint of  a   channel-independent velocity.  In  Fig.~\ref{fig:spectral-functions}, we use  the  ALF~\cite{ALF_v2}  implementation of  the  stochastic maximum entropy methods~\cite{beach04}  to compute  the spectral  functions in the single  particle,
\begin{equation} 
    A_c(\boldsymbol{k},\omega) =    \pi \sum_{n,\lambda}  \left| \langle n | \mathbf{\hat{c}}^{\dagger}_{\boldsymbol{k},\lambda} | 0 \rangle \right|^2 
    \delta(  E_n - E_0 - \omega),
\end{equation} 
particle-hole,  
\begin{equation} 
    \chi^s(\boldsymbol{k},\omega) =   \pi  \sum_{n,\lambda}  \left| \langle n | \hat{\vec{s}}_{\boldsymbol{k},\lambda} | 0 \rangle \right|^2 
    \delta(  E_n - E_0 - \omega),
\end{equation} 
and particle-particle,
\begin{equation} 
    \chi^c(\boldsymbol{k},\omega) =    \pi \sum_{n}  \left| \langle n | {\hat{\Delta}}^{\dagger}_{\boldsymbol{k}} | 0 \rangle \right|^2 
    \delta(  E_n - E_0 - \omega),
\end{equation}
channels,
where $|n\rangle$ and $E_n$, $n=0,1,2,\dots,$ correspond to the energy eigenstates and eigenvalues, respectively.
Here,  we  consider    $J \approx J_\text{c2} = 1.057(10)$.   As  apparent  from Fig.~\ref{fig:spectral-functions},
all  quantities  show  a linear  dispersion relation at low energy. The single-particle spectral function is gapless at the $\mathbf K$ and $\mathbf K'$ points in the Brillouin zone, while the particle-particle and particle-hole spectral functions are gapless at the  $\boldsymbol{\Gamma}$ point.  
The gapless excitations in the particle-particle and particle-hole channels can be understood as collective modes associated with the SO(3) and U(1) order parameters, which characterize the neighboring long-range-ordered phases.
Importantly,  the  velocity  in all  three aforementioned channels are  comparable,  such  that the data  support    emergent 
Lorentz symmetry   characterized  by a  single ``velocity of light.''

\section{Finite-temperature results}
\label{sec:thermodynamics}

Further insight into the nature of the SO(3)-U(1) transition can be obtained by studying the finite-temperature properties above the potential quantum critical point.
Since the interlayer-coherent insulating or $s$-wave superconducting state at $J > J_\mathrm{c2}$ breaks U(1) charge and $\mathds{Z}_2$ PPH symmetries only, we expect that the gapped phase extends to finite temperatures up to a critical temperature $T_\mathrm{c}(J)$.
By contrast, the SO(3)-spin-symmetry-broken state for $J<J_\mathrm{c2}$, is expected to destabilize at arbitrary small temperatures as a consequence of the Mermin-Wagner theorem.

In order to investigate the finite-temperature properties, we compute the uniform susceptibilities~\cite{hohenadler22}
\begin{align}
\chi_\text{uni}^\mathrm{SO(3)} = \frac{\beta}{L^2} \left( \langle \hat{\vec{S}} \cdot \hat{\vec{S}} \rangle - \langle \hat{\vec{S}} \rangle \cdot \langle \hat{\vec{S}} \rangle \right)
\end{align}
measuring SO(3) spin fluctuations, with the total-spin operator $\hat{\vec{S}}  = (\hat{S}^\alpha) = \sum_{i} {\hat{c}}_{i,\sigma,\lambda}^{\dagger}K^\alpha_{\sigma\sigma'} {\hat{c}}_{i,\sigma',\lambda}$, which generates SO(3) rotations, and
\begin{align}
\chi_\text{uni}^\mathrm{U(1)} = \frac{\beta}{L^2} \left( \langle \hat{\rho} \hat{\rho} \rangle - \langle \hat{\rho} \rangle \langle \hat{\rho} \rangle \right)
\end{align}
measuring U(1) charge fluctuations, with the total-charge operator $\hat{\rho} = \sum_i {\hat{c}}^\dagger_{i,\sigma,\lambda} {\hat{c}}_{i,\sigma,\lambda}$, which generates U(1) rotations.

\begin{figure}[tb]
\includegraphics[width=\columnwidth]{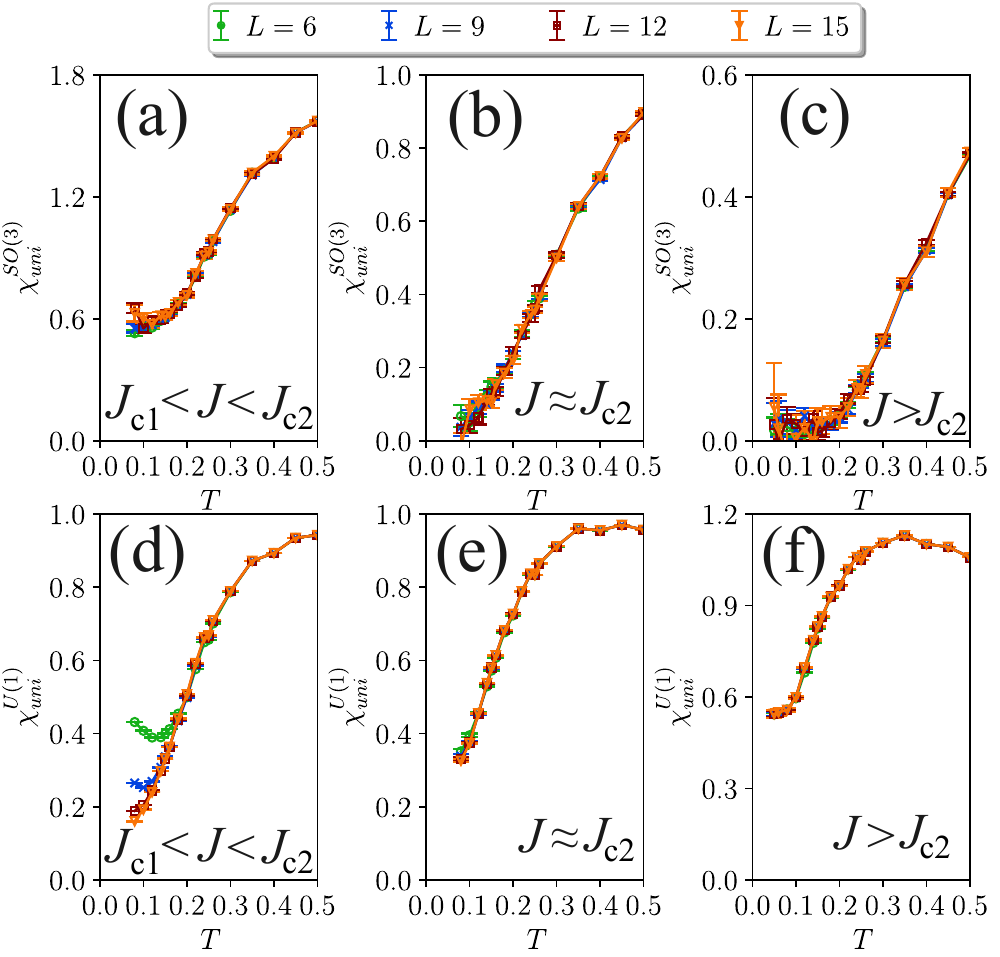}
\caption{%
(a)~Uniform SO(3) susceptibility $\chi_\text{uni}^\mathrm{SO(3)}$ as function of temperature for a representative fixed value of $J=0.80$ below the SO(3)-U(1) transition point.
(b)~Same as (a), but for a value of $J=1.06$ close to the SO(3)-U(1) transition point.
(c)~Same as (a), but for a value of $J=1.30$ above the SO(3)-U(1) transition point.
(d)~Uniform SO(3) susceptibility $\chi_\text{uni}^\mathrm{U(1)}$ as function of temperature for a representative fixed value of $J=0.80$ below the SO(3)-U(1) transition point.
(e)~Same as (d), but for a value of $J=1.06$ close to the SO(3)-U(1) transition point.
(f)~Same as (d), but for a value of $J=1.30$ above the SO(3)-U(1) transition point.
\label{fig:uni_sus}}
\end{figure}

The results for the uniform susceptibilities as function of temperature are shown for three different values below, near, and above, respectively, the SO(3)-U(1) transition point in Fig.~\ref{fig:uni_sus}.
For strong coupling $J > J_\mathrm{c2}$ above the interlayer-coherent insulating or $s$-wave superconducting ground state, the SO(3) susceptibility $\chi_\text{uni}^\mathrm{SO(3)}$ is strongly suppressed at low temperatures, in agreement with the expectation of an exponential decay, see Fig.~\ref{fig:uni_sus}(c). This reflects the gapped nature of the SO(3) spin spectral function in this phase.
By contrast, the U(1) susceptibility $\chi_\text{uni}^\mathrm{U(1)}$ approaches a finite value in the low-temperature limit, see Fig.~\ref{fig:uni_sus}(f), reflecting the low-energy spectral weight in the U(1) charge spectral function in this phase.
For intermediate coupling $J_\mathrm{c1} < J < J_\mathrm{c2}$ above the SO(3)-spin-symmetry-broken ground state, the SO(3) susceptibility $\chi_\text{uni}^\mathrm{SO(3)}$ approaches a finite value in the low-temperature limit [Fig.~\ref{fig:uni_sus}(a)], indicating the onset of short-range  order in the SO(3) spin sector. The U(1) susceptibility $\chi_\text{uni}^\mathrm{U(1)}$ is consistent with a linear behavior as function of temperature $T$ at low $T$, reflecting the semimetallic behavior of the underlying SO(3) ground state [Fig.~\ref{fig:uni_sus}(d)].
At finite temperatures above a quantum critical point, we expect that the uniform susceptibilities scale with temperature as $\chi_\text{uni} \propto T^{2/z-1}$, where $z$ corresponds to the dynamical critical exponent~\cite{hohenadler22}. Assuming $z=1$ leads to a linear-in-$T$ behavior.
In the vicinity of the SO(3)-U(1) transition point at $J_\mathrm{c2}$, our data for both $\chi_\text{uni}^\mathrm{SO(3)}$ and $\chi_\text{uni}^\mathrm{U(1)}$ are indeed consistent with a linear-in-$T$ dependence in the low-temperature limit, see Figs.~\ref{fig:uni_sus}(b) and \ref{fig:uni_sus}(e).
This again confirms our interpretation of a continuous or weakly-first-order SO(3)-U(1) transition.

\begin{figure}[tb]
\begin{centering}
\includegraphics[width=\columnwidth]{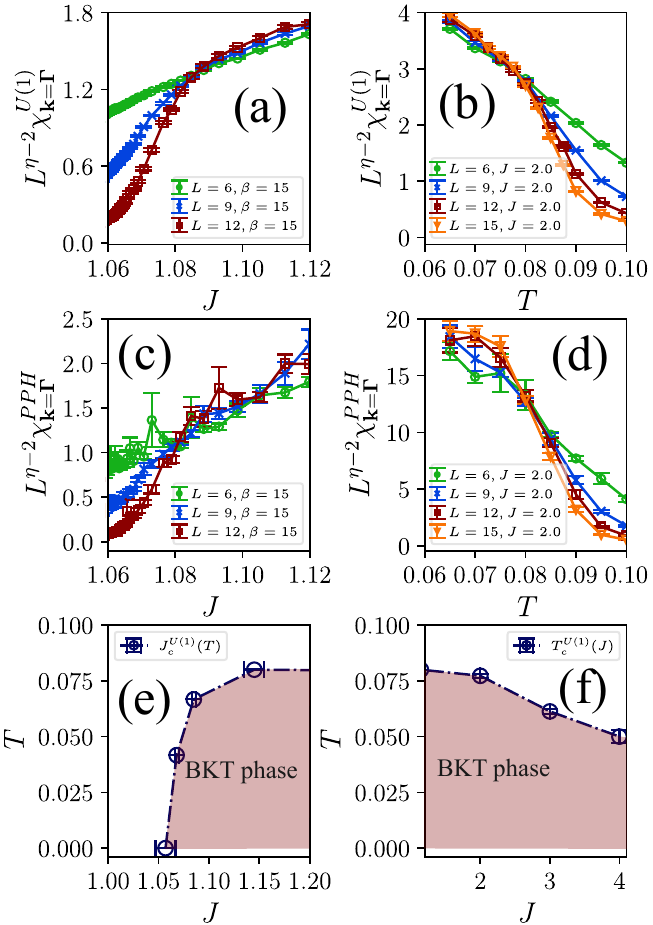}
\par\end{centering}
\caption{%
(a)~Rescaled U(1) susceptibility $L^{\eta-2}\chi^\mathrm{U(1)}_{\boldsymbol{k} = \boldsymbol{\Gamma}}$ for $\eta = 0.25$ as function of coupling $J$ for different system sizes $L$ and a representative fixed inverse temperature $\beta = 15$. The crossing point indicates the phase boundary $J_\mathrm{c}^\mathrm{U(1)}(T)$ at finite temperature $T = 1/\beta$.
(b)~Rescaled U(1) susceptibility $L^{\eta-2}\chi^\mathrm{U(1)}_{\boldsymbol{k} = \boldsymbol{\Gamma}}$ for $\eta = 0.25$ as function of temperature $T$ for different system sizes $L$ and a representative large value of the coupling $J = 2$ well above the SO(3)-U(1) transition. The crossing point indicates the finite-temperature phase boundary $T_\mathrm{c}^\mathrm{U(1)}(J)$.
(c,d)~Same as (a,b), but for the rescaled PPH susceptibility $L^{\eta-2}\chi^\mathrm{PPH}_{\boldsymbol{k} = \boldsymbol{\Gamma}}$.
(e)~Finite-temperature phase diagram in the vicinity of the SO(3)-U(1) transition from crossings of rescaled U(1) susceptibility as function of $J$. Red region refers to phase characterized by algebraic order in the U(1)-charge sector (``BKT phase'').
(f)~Same as (c), but at large coupling $J$ well above the SO(3)-U(1) transition from crossings of rescaled U(1) susceptibility as function of $T$.
\label{fig:ft_phase}}
\end{figure}

In order to characterize the finite-temperature phase boundary $T_\mathrm{c}^\text{U(1)}(J)$, below which U(1) symmetry is spontaneously broken  for $J> J_\mathrm{c2}$, we also compute the momentum-resolved U(1) susceptibility
\begin{align}
\chi^\text{U(1)}_{\boldsymbol{k}}=\int \rmd \tau \mathcal S_\text{U(1)} (\boldsymbol{k},\tau),
\end{align}
where $\mathcal S_\text{U(1)}$ is the U(1) structure factor defined in Eq.~\eqref{eq:U(1)-structure-factor}.
We expect the low-temperature phase at $T<T_\mathrm{c}^\text{U(1)}(J)$ to be bounded by a Berezenskii-Kosterlitz-Thouless (BKT) transition~\cite{berezinskii71, kosterlitz73, kosterlitz74}, associated with the U(1)-charge-symmetry-broken ground state at $J>J_\mathrm{c2}$.
The low-temperature phase is characterized by algebraic order that scales as $r^{-\eta}$ as function of distance $r$, with temperature-dependent anomalous dimension $\eta = \eta(T)$ for $T \leq T_\mathrm{c}(J)$.
Right at the BKT transition, the anomalous dimension becomes $\eta(T_\mathrm{c}) = 0.25$.
As a consequence, the BKT transition temperature can therefore be located by a crossing-point analysis of the rescaled U(1) susceptibility $L^{\eta-2}\chi^\text{U(1)}_{\boldsymbol{k} = \boldsymbol{\Gamma}}$ with $\eta=0.25$.
Instead of scanning as function of temperature for fixed $J$, we may also use the coupling $J$ as tuning parameter for the transition at fixed finite temperature $T$. This procedure turns out to be particularly useful in the vicinity of the SO(3)-U(1) quantum phase transition, in the vicinity of which the finite-temperature phase boundary in the plane spanned by $J$ and $T$ is very steep.
The rescaled U(1) susceptibility $L^{\eta-2}\chi^\mathrm{U(1)}_{\boldsymbol{k} = \boldsymbol{\Gamma}}$ as function of $J$ for different system sizes $L$ and a representative fixed temperature is shown in Fig.~\ref{fig:ft_phase}(a). The corresponding crossing points $J_\mathrm{c}^\mathrm{U(1)}(T)$ are shown in the finite-temperature phase diagram for $J$ near $J_\mathrm{c2}$ in Fig.~\ref{fig:ft_phase}(e).
For larger values of $J$, the phase boundary in the plane spanned by $J$ and $T$ is rather flat, and it is more convenient to use the temperature as tuning parameter at fixed coupling $J$.
The rescaled U(1) susceptibility $L^{\eta-2}\chi^\text{U(1)}_{\boldsymbol{k} = \boldsymbol{\Gamma}}$ as function of $T$ for different system sizes $L$ and a representative fixed large value of $J$ is shown in Fig.~\ref{fig:ft_phase}(b). The corresponding crossing points $T_\mathrm{c}^\mathrm{U(1)}(J)$ are shown in the finite-temperature phase diagram for large $J$ well above $J_\mathrm{c2}$ in Fig.~\ref{fig:ft_phase}(f).

In the low-temperature phase for $T<T_\mathrm{c}^\text{U(1)}$, the system spontaneously selects an $s$-wave superconducting or interlayer-coherent insulating state as ground state. This implies that $\mathds{Z}_2$ PPH symmetry is broken in this phase as well.
As a consequence, a lower bound for the critical temperature $T_\mathrm{c}^\text{PPH}$, marking the melting of PPH order, is given by the U(1) critical temperature $T_\mathrm{c}^\text{U(1)}$, i.e., $T_\mathrm{c}^\text{PPH} \geq T_\mathrm{c}^\text{U(1)}$. 
An interesting question is whether the two critical temperatures coincide or differ. The latter scenario would imply an intermediate vestigial phase for $T_\mathrm{c}^\text{U(1)} < T < T_\mathrm{c}^\text{PPH}$, in which PPH symmetry is spontaneously broken, but both U(1) global-charge and U(1) layer-charge symmetries remain intact. Such vestigial orders have previously been discussed in a variety of two-dimensional models~\cite{svistunov15, fernandes19, francini24}.
In order to characterize the melting of PPH order, we measure the PPH susceptibility
\begin{align}
\chi^\text{PPH}_{\boldsymbol{k}}=\int \rmd \tau \mathcal S_\text{PPH} (\boldsymbol{k},\tau),
\end{align}
where $\mathcal S_\text{PPH}$ is the PPH structure factor defined analogously to Eq.~\eqref{eq:U(1)-structure-factor}, using the parity operator $\hat P_i$ instead of the interlayer coherence operator $\hat n_i$.
From the symmetry of the order parameter, we expect the PPH transition, if continuous, to fall into the 2D Ising universality class.
In Fig.~\ref{fig:ft_phase}(c), we therefore show the rescaled PPH susceptibility $L^{\eta-2}\chi^\mathrm{U(1)}_{\boldsymbol{k} = \boldsymbol{\Gamma}}$ as function of $J$ for different system sizes $L$ and a representative fixed temperature [same temperature as those of the U(1) susceptibility shown in Fig.~\ref{fig:ft_phase}(a)], using the Ising exponent $\eta = 0.25$.
Similarly, we show in Fig.~\ref{fig:ft_phase}(d) the rescaled PPH susceptibility as function of $T$ for different system sizes $L$ and a representative fixed large value of $J$ [same coupling as those of the U(1) susceptibility shown in Fig.~\ref{fig:ft_phase}(b)] using $\eta=0.25$.
While the data clearly indicate the PPH order at sufficiently large $J$ and low $T$, they are too noisy to clearly identify whether the critical temperature $T_\mathrm{c}^\text{PPH}$ lies above or right at $T_\mathrm{c}^\text{U(1)}$.
This is also due to the fact that vestigial orders are typically realized in only small temperature windows above the corresponding primary orders~\cite{svistunov15, fernandes19, francini24}. The presence of a tiny vestigial phase above $T_\mathrm{c}^\text{U(1)}$ can therefore at present not be excluded from our data. Exploring this possibility deserves further investigation.

\section{Conclusions}
\label{sec:conclusions}

In this work, we have characterized, by means of large-scale determinant QMC simulations, the metallic and deconfined quantum phase transitions recently discovered in a bilayer honeycomb model with an SO(3)-symmetric spin-density interaction in terms of their quantum critical and finite-temperature properties.
In comparison with our initial exploratory study of this model~\cite{liu22}, we have employed a Hermitian Trotter decomposition, which significantly improves convergence properties with respect to the limit of small Trotter time steps.
Furthermore, we have employed a microscopic implementation that preserves the model's partial particle-hole symmetry explicitly, maintaining the degeneracy of the interlayer-coherent insulating and $s$-wave superconducting ground states in the fully gapped phase stabilized at strong coupling already on finite lattice sizes.

These advances have lead to improved estimates for the critical exponents characterizing the Gross-Neveu-SO(3) quantum critical point at $J_\mathrm{c1}$. From the data-collapse analysis, we have obtained $1/\nu = 0.86(8)$ for the correlation-length exponent, $\eta_\phi = 0.73(2)$ for the order-parameter anomalous dimension, and $\eta_\psi = 0.078(8)$ for the fermion anomalous dimensions. These results are consistent with ealier field-theoretical estimates~\cite{ray21} within the systematics uncertainties of both the numerical and field-theoretical approaches.

We have also provided a further characterization of the putative deconfined metallic quantum critical point at $J_\mathrm{c2}$.
In particular, the spectral functions in the single-particle, particle-hole, and particle-particle channel indicate gapless excitations with a unique ``velocity of light,'' supporting the emergence of Lorentz symmetry at $J_\mathrm{c2}$.
We have also computed the finite-temperature phase diagram of the model and show that the boundary of the low-temperature phase vanishes continuously upon approaching $J_\mathrm{c2}$, in agreement with the expectation for a continuous or weak-first-order transition.

Our microscopic model features an explicit $\mathrm{SO(3)} \times \mathrm{U}(1) \times \mathrm{U}_\mathrm{L}(1) \times \mathds{Z}_2$ symmetry in the spin, total-charge, layer-charge, and partial-particle-hole sectors.
For the future, it would be interesting to investigate the possibility of an emergent higher symmetry at the SO(3)-U(1) transition. To this end, the decay of dynamical correlation functions of higher-symmetry generators should be studied.
This could help to develop a field-theoretical understanding of the putative deconfined metallic transition, e.g., via a Wess-Zumino-Witten theory coupled to fermionic degrees of freedom. 

Furthermore, it would be interesting to study the effects of explicit symmetry breaking, e.g., in the SO(3) spin sector. The Gross-Neveu-SO(3) transition at $J_\mathrm{c1}$ is expected to become a multicritical point featuring emergent SO(3) symmetry within the enlarged parameter space, and the SO(3)-spin-ordered semimetal phase at intermediate coupling $J_\mathrm{c1} < J < J_\mathrm{c2}$ is destabilized in favor of a SO(2)- or $\mathds{Z}_2$-spin-ordered phase by the symmetry-breaking perturbations~\cite{fornoville24}. 
Since the U(1)-ordered phase at strong coupling $J>J_\mathrm{c2}$ is gapped, one might expect that it is stable upon adding small symmetry-breaking perturbations in the spin sector. If that is correct, it would be interesting to study the nature of the transition between the SO(2)- or $\mathds{Z}_2$-spin-ordered phases at intermediate coupling and the U(1)-ordered phase at strong coupling. This should be expected to also lead to further insights into the nature of the putative deconfined metallic quantum critical point at $J_\mathrm{c2}$.

A recent field-theoretical analysis~\cite{ray24} proposed the existence of another quantum critical point in a new universality class, which may be reached within the parameter space of our model by adding to the Hamiltonian an interaction of the form
\begin{align}
 \hat{H}_{J'} & = -J' \sum_{i,\alpha<\beta} \left(\hat{c}_{i,\sigma,\lambda}^{\dagger}Q_{\sigma\sigma^{\prime}}^{\alpha\beta}\hat{c}_{i,\sigma^{\prime},\lambda}\right)^{2},
\end{align}
with the real $3\times 3$ matrices $Q^{\alpha\beta} = \frac12 \{K^\alpha, K^\beta\} - \frac23 \delta^{\alpha\beta}$.
It would be interesting to investigate this conjecture numerically. As the vertex matrix associated with the above interaction is real, the model can be simulated using QMC simulations without a fermion sign problem for positve values  of  $J'$.


\begin{acknowledgments}
We thank Shouryya Ray for discussions and collaborations on related problems and
Subhro Bhattacharjee for discussions on spin-orbit coupled models showing similar phases.
The authors gratefully acknowledge the Gauss Centre for Supercomputing e.V.\ (www.gauss-centre.eu) for funding this project by providing computing time on the GCS Supercomputer SUPERMUC-NG at Leibniz Supercomputing Centre (www.lrz.de).
This research has been supported by the Deutsche Forschungsgemeinschaft through the W\"urzburg-Dresden Cluster of Excellence on Complexity and Topology in Quantum Matter -- \textit{ct.qmat} (EXC 2147, Project No.\ 390858490), SFB 1170 on Topological and Correlated Electronics at Surfaces and Interfaces (Project No.\  258499086), SFB 1143 on Correlated Magnetism (Project No.\ 247310070), and the Emmy Noether Program (JA2306/4-1, Project No.\ 411750675).
\end{acknowledgments}

\bibliographystyle{longapsrev4-2}
\bibliography{SO3}


\end{document}